\documentclass[sigconf]{acmart}

\AtBeginDocument{%
  }

\usepackage{tcolorbox}

\usepackage{bbding}
\usepackage{subfigure}
\usepackage[boxed,ruled,lined]{algorithm2e}
\usepackage{multirow}
\usepackage{makecell}

\usepackage{arydshln} 
\usepackage{enumerate}
\usepackage{enumitem}
\usepackage{makecell}

\usepackage{booktabs}
\usepackage{xspace}
\usepackage{color}
\usepackage{xcolor}
\usepackage{subfigure}
\usepackage{pifont}

\newcommand{\srcandrec}{S\&R\xspace}
\newcommand{\ourname}{LCR-SER\xspace}

\setcopyright{acmlicensed}
\copyrightyear{2018}
\acmYear{2018}
\acmDOI{XXXXXXX.XXXXXXX}
\acmConference[Conference acronym 'XX]{Make sure to enter the correct
  conference title from your rights confirmation email}{June 03--05,
  2018}{Woodstock, NY}
\acmISBN{978-1-4503-XXXX-X/2018/06}

\begin{document}

\title{Bridging Search and Recommendation through \\ Latent Cross Reasoning}

\author{Teng Shi}
\author{Weicong Qin}
\affiliation{
\institution{\mbox{Gaoling School of Artificial Intelligence}\\Renmin University of China}
  \city{Beijing}
  \country{China}
}
\email{{shiteng,qwc}@ruc.edu.cn}

\author{Weijie Yu}
\affiliation{
\institution{School of Information Technology and Management\\University of International Business and Economics}
  \city{Beijing}
  \country{China}
}
\email{yu@uibe.edu.cn}

\author{Xiao Zhang}
\affiliation{
\institution{\mbox{Gaoling School of Artificial Intelligence}\\Renmin University of China}
  \city{Beijing}
  \country{China}
}
\email{zhangx89@ruc.edu.cn}

\author{Ming He}
\affiliation{
\institution{AI Lab at Lenovo Research}
  \city{Beijing}
  \country{China}
}
\email{heming01@foxmail.com}

\author{Jianping Fan}
\affiliation{
\institution{AI Lab at Lenovo Research}
  \city{Beijing}
  \country{China}
}
\email{jfan1@lenovo.com}

\author{Jun Xu}
\affiliation{
\institution{\mbox{Gaoling School of Artificial Intelligence}\\Renmin University of China}
  \city{Beijing}
  \country{China}
}
\email{junxu@ruc.edu.cn}

\renewcommand{\shortauthors}{Teng Shi et al.}

\begin{abstract}
Search and recommendation (\textbf{\srcandrec}) are fundamental components of modern online platforms, yet effectively leveraging search behaviors to improve recommendation remains a challenging problem. User search histories often contain noisy or irrelevant signals that can even degrade recommendation performance, while existing approaches typically encode \srcandrec histories either jointly or separately without explicitly identifying which search behaviors are truly useful.
Inspired by the human decision-making process, where one first identifies recommendation intent and then reasons about relevant evidence, we design a latent cross reasoning framework that first encodes user \srcandrec histories to capture global interests and then iteratively reasons over search behaviors to extract signals beneficial for recommendation. Contrastive learning is employed to align latent reasoning states with target items, and reinforcement learning is further introduced to directly optimize ranking performance.
Extensive experiments on public benchmarks demonstrate consistent improvements over strong baselines, validating the importance of reasoning in enhancing search-aware recommendation.

\end{abstract}



\maketitle

\section{Introduction}

Search and recommendation (\textbf{\srcandrec}) have become core components of many commercial applications. Users rely on both to fulfill different types of information needs, which in turn reflect diverse aspects of their interests. 
These two types of behaviors are complementary, as user preferences observed in one can be leveraged to enhance the effectiveness of the other.

To exploit this complementarity, many existing studies have explored the joint modeling of \srcandrec~\cite{SESRec,xie2024unifiedssr,UniSAR}.
These methods can be divided into two categories:
(1)~Unified Encoder, which encodes \srcandrec histories as a single sequence by interleaving them in chronological order~\cite{USER,UniSAR,shi2025unified};
(2) Dual Encoder, which treats \srcandrec histories as two separate sequences and encodes them independently~\cite{SESRec,xie2024unifiedssr}.

\begin{figure}[t]
    \centering
    \includegraphics[width=0.75\columnwidth]{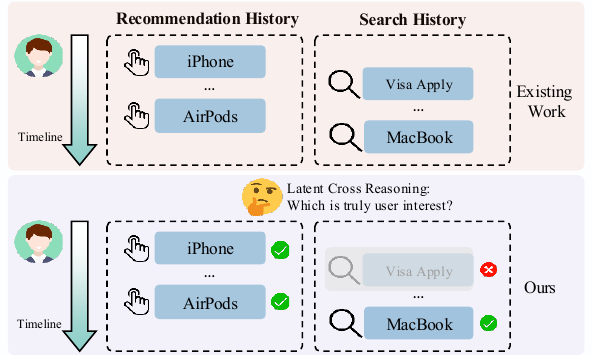}
    \caption{
    An example showing that search history may contain irrelevant queries (e.g., ``Visa Apply'') unrelated to recommendation interests (e.g., electronics). While existing methods treat all queries equally, our approach reasons to extract relevant signals for improved modeling.
    }
    \label{fig:intro_example}
\end{figure}

While existing methods have achieved promising results, they typically adopt either unified or dual encoding strategies that directly treat the encoded representations of user \srcandrec histories as signals for prediction. However, these approaches do not explicitly distinguish which parts of the search interest are truly beneficial for modeling recommendation preferences. In practice, not all search behaviors are helpful, and some may even introduce noise. As illustrated in Figure~\ref{fig:intro_example}, a user might search for ``Visa Apply'',
while their recommendation history is dominated by electronic products. Naively incorporating such unrelated search queries can harm recommendation performance.

\begin{figure}[t]
    \centering
    \includegraphics[width=0.7\columnwidth]{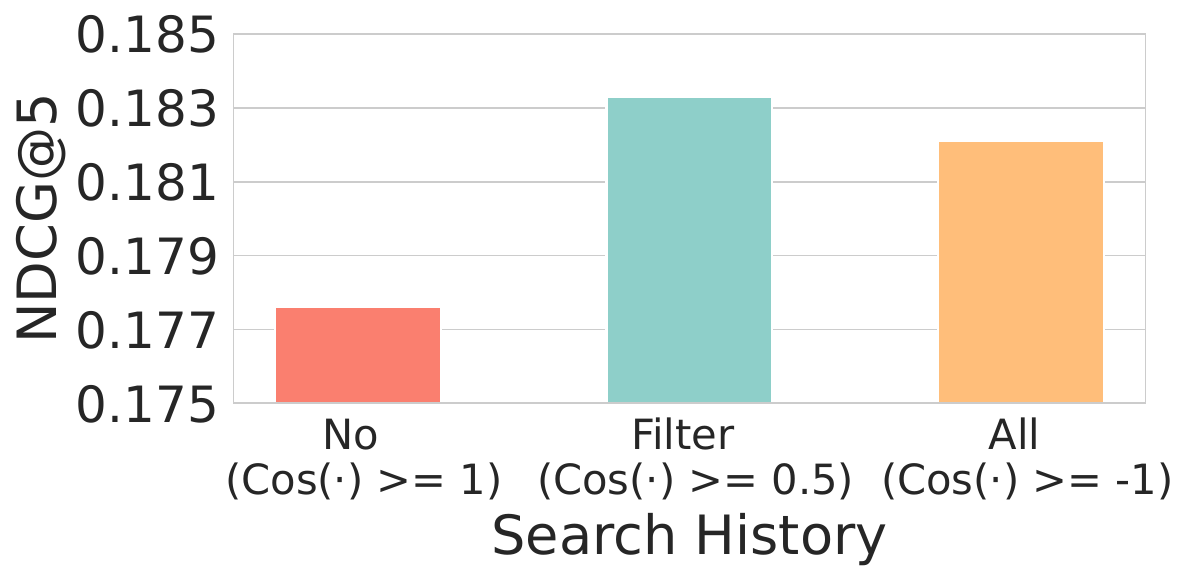}
    \caption{
    Empirical analysis of the state-of-the-art joint \srcandrec model UniSAR~\cite{UniSAR} on KuaiSAR-Small~\cite{sun2023kuaisar} dataset: each search behavior is compared with the average embedding of the recommendation history via cosine similarity. Search behaviors with similarity above a threshold are retained, where 1 removes all, 0.5 retains only the most relevant, and -1 preserves all.
    }
    \label{fig:intro_analysis}
\end{figure}

To further validate our motivation, we conduct an empirical analysis using a simple filtering heuristic based on the cosine similarity between \srcandrec behaviors (Figure~\ref{fig:intro_analysis}).
We observe that removing all search history degrades recommendation performance compared to including it, 
confirming the overall utility of search signals. Moreover, retaining only the most relevant search behaviors outperforms using the entire history, validating our motivation that selectively leveraging search signals is crucial.
However, this empirical approach relies on manually defined thresholds and cannot adaptively select relevant search histories.

In contrast, human decision-making often involves an initial holistic understanding of past behaviors, followed by iterative reasoning to identify which information is relevant to the current task. Humans naturally refine their decisions step by step, using feedback from the reasoning process. However, current encoding-based models lack this iterative refinement mechanism and instead rely on one-pass encoding without selective reasoning.

To bridge this gap, we draw inspiration from the emerging paradigm of latent reasoning in large language models (LLMs)~\cite{zhu2025survey,hao2024training}. Latent reasoning enables models to refine their internal representations by feeding hidden states back into the model across multiple steps, offering greater efficiency compared to explicit reasoning~\cite{guo2025deepseek}. Despite its success in natural language tasks, its application in recommendation remains limited~\cite{liu2025lares,tang2025think}. This challenge is especially pronounced in search-enhanced recommendation scenarios, where users exhibit dual behavior patterns: search interactions involve textual queries, while recommendation interactions are item-centric. The heterogeneity and complexity of these behaviors pose additional challenges for applying latent reasoning effectively.

To address the above challenges, we propose \textbf{\ourname}, a \textbf{L}atent \textbf{C}ross \textbf{R}easoning framework for \textbf{S}earch \textbf{E}nhanced 
\textbf{R}ecommendation task.
We first encode the user's \srcandrec histories using separate encoders. Then, we perform latent reasoning by feeding the final hidden states back into their respective encoders. This enables the model to revisit and refine representations, progressively aligning them with the user's underlying interests.
To identify useful search signals for recommendation, we introduce cross attention during reasoning, allowing the recommendation encoder to incorporate information from the search encoder and vice versa. This encourages mutual interest alignment across behaviors.
To further enhance relevance-aware reasoning, we apply contrastive learning to bring cross-attentive representations closer to the target item embedding compared to non-attentive ones.
Finally, we adopt GRPO-based reinforcement learning (RL)~\cite{guo2025deepseek} 
to optimize the reasoning process. Ranking metrics serve as the reward signal, and Gaussian noise is injected into the initial reasoning state to encourage exploration of diverse reasoning~paths.

In summary, this paper makes the following contributions:
\begin{itemize}[leftmargin=1em]
    \item We identify a key limitation in existing joint modeling of \srcandrec: the lack of effective mechanisms to extract search interests that are truly beneficial for recommendation.
    \item We propose a novel method, \ourname, which employs latent cross reasoning to infer user interests more effectively across \srcandrec. We further enhance the model with GRPO-based training for improved reasoning.
    \item Extensive experiments on public datasets demonstrate that \ourname consistently outperforms both traditional and search-enhanced recommendation baselines.
\end{itemize}

\section{Related Work}
\subsection{Search Enhanced Recommendation}
Search~\cite{zhang2025trigger3,qin2025similarity,shi2025retrieval,shen2024survey,qin2024explicitly,qin2025maps,qin2025uncertainty,zhang2025syler,sun2024logic,zhang2024citalaw} and recommendation~\cite{zhang2024model,zhang2024modeling,zhang2024qagcf,zhang2025test,zhang2024reinforcing,dai2024modeling,shen2024generating,SAQRec,qin2024enhancing} (\textbf{\srcandrec}) are central to modern online services.
They capture complementary aspects of user intent, enabling behaviors in one to enhance the effectiveness of the other.
Search Enhanced Recommendation
has attracted growing interest, with recent works increasingly leveraging search data to improve recommendation models~\cite{IV4REC, wang2024enhancing, SRJgraph, zhang2024unified, shi2025unified,zhao2025unifying,penha2024bridging}. JSR~\cite{JSR, JSR2} jointly trains \srcandrec models with a shared loss, while USER~\cite{USER} encodes both behaviors using a Transformer. SESRec~\cite{SESRec} applies contrastive learning to distinguish similar and dissimilar interests across \srcandrec. UnifiedSSR~\cite{xie2024unifiedssr} adopts a dual-branch network to separately encode query and product histories. UniSAR~\cite{UniSAR} models user transitions between \srcandrec using Transformers with distinct masking.
Unlike existing work, we enhance recommendation with search data through latent cross~reasoning.

\subsection{Latent Reasoning}
Latent reasoning 
has recently gained attention in large language models (LLMs)~\cite{zhu2025survey,hao2024training,geiping2025scaling}. By feeding the final hidden state back into the model, it enables reasoning in the hidden space, offering greater efficiency compared to explicit reasoning~\cite{guo2025deepseek,yang2025qwen3}. Recent work has also applied latent reasoning to recommender systems~\cite{tang2025think,liu2025lares,zhang2025reinforced}, enhancing user interest modeling and performance. This paper explores its use in search-enhanced recommendation to better integrate user interests from both \srcandrec behaviors.

\section{Method}
\begin{figure*}[t]
    \centering
    \includegraphics[width=1.0\linewidth]{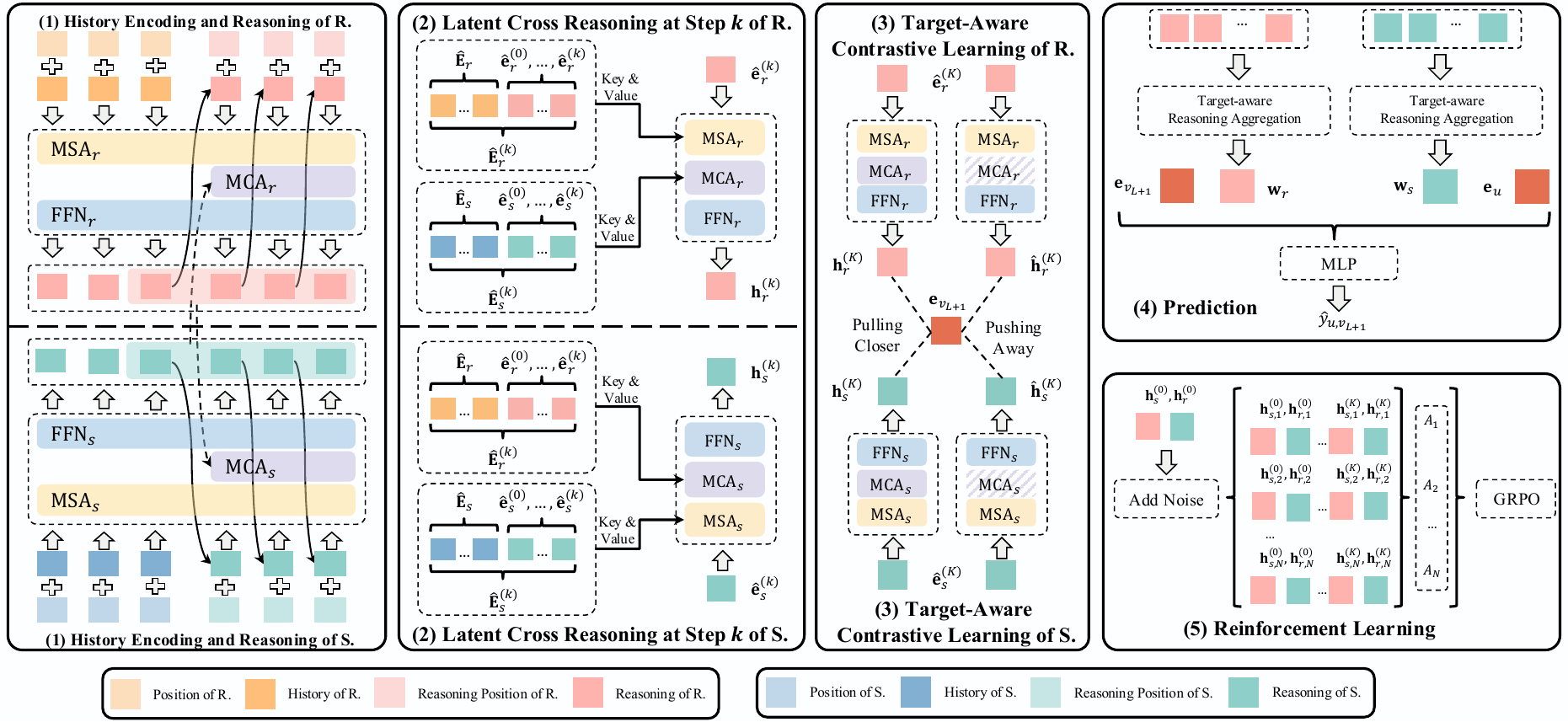}
    \caption{
    Overall framework of \ourname:
    (1)~\srcandrec histories are independently encoded.
    (2)~Latent cross reasoning iteratively refines \srcandrec representations via cross-history attention.
    (3)~Contrastive learning aligns reasoning hidden states with the target item to guide reasoning.
    (4)~Refined representations are aggregated for the final prediction.
    (5)~Reinforcement learning further optimizes ranking performance.
    ``S.'' and ``R.'' denote search and recommendation, respectively.
    }
\label{fig:method}
\end{figure*}

This section introduces our method \ourname (Figure~\ref{fig:method}). We first formulate the search-enhanced recommendation task, then present the latent cross reasoning architecture. We also describe the supervised pre-training with contrastive and recommendation losses, followed by a reinforcement learning strategy to enhance reasoning.

\subsection{Problem Formulation}
Let $\mathcal{U}$, $\mathcal{V}$, and $\mathcal{Q}$ denote the sets of users, items, and queries. Each user $u \in \mathcal{U}$ has a recommendation history $\mathcal{R}_u = [v_1, \dots, v_{L_r}]$ and a search history $\mathcal{S}_u = [(q_1, \mathcal{V}_{q_1}), \dots, (q_{L_s}, \mathcal{V}_{q_{L_s}})]$, where $v_t \in \mathcal{V}$ is the $t$-th clicked item, and each $q_t \in \mathcal{Q}$ is a user-issued query followed by a set of $n_t$ clicked items $\mathcal{V}_{q_t} = \{v^{(t)}_1, \dots, v^{(t)}_{n_t}\}$. The total number of interactions is $L = L_r + L_s$.
Our goal is to learn a model $\Phi$ that predicts the next item $v_{L+1}$ using both $\mathcal{R}_u$ and~$\mathcal{S}_u$.

\subsection{Architecture}
This section outlines the architecture of \ourname. We first encode \srcandrec histories with Transformer encoders. A latent cross reasoning module then iteratively integrates information across both behaviors to refine user representations. Finally, we aggregate the multi-step hidden states based on their similarity to the target for final prediction.

\subsubsection{Embedding Layer}
We maintain three separate embedding matrices to represent users, items, and query words:
$\mathbf{E}_{\mathcal{U}} \in \mathbb{R}^{|\mathcal{U}| \times d}$,
$\mathbf{E}_{\mathcal{V}} \in \mathbb{R}^{|\mathcal{V}| \times d}$, and
$\mathbf{E}_{\mathcal{W}} \in \mathbb{R}^{|\mathcal{W}| \times d}$,
where $\mathcal{W}$ denotes the vocabulary of all words appearing in user queries, and $d$ is the embedding dimension.
Given a user $u$ and an item $v$, their corresponding embeddings $\mathbf{e}_u \in \mathbb{R}^{d}$ and $\mathbf{e}_v \in \mathbb{R}^{d}$ are obtained via standard table lookup.
For a query $q = \{ w_1, w_2, \dots, w_{|q|} \} \subseteq \mathcal{W}$ consisting of a sequence of words, we follow prior works~\cite{SESRec,UniSAR} and represent it by averaging the embeddings of its constituent words:
$\mathbf{e}_q = \frac{1}{|q|} \sum_{i=1}^{|q|} \mathbf{e}_{w_i}$,
where $\mathbf{e}_{w_i} \in \mathbb{R}^{d}$ is the embedding of the $i$-th word in the query.

\subsubsection{History Encoding}
We represent the user's behavioral history by separately encoding \srcandrec interactions, based on embeddings retrieved via standard lookup operations.
For the recommendation history $\mathcal{R}_u$, we construct an embedding matrix by stacking the embeddings of the corresponding items:
$\mathbf{E}_{r}=[\mathbf{e}_{v_1},
\ldots,\mathbf{e}_{v_{L_r}}]^{\mathsf{T}} \in \mathbb{R}^{L_r \times d}.$
For the search history $\mathcal{S}_u$,
we represent each interaction by summing the query embedding and the average embedding of the clicked items:
$\mathbf{E}_{s}=[
\mathbf{e}_{q_1} + \mathrm{M}(\mathcal{V}_{q_1}),
\ldots,\mathbf{e}_{q_{L_s}} + \mathrm{M}(\mathcal{V}_{q_{L_s}})]^{\mathsf{T}} \in \mathbb{R}^{L_s \times d},$
where 
$\mathrm{M}(\mathcal{V}_{q_t}) = \frac{1}{n_t} \sum_{i=1}^{n_t} \mathbf{e}_{v^{(t)}_{i}}$ denotes the mean-pooled embedding of items clicked in response to query $q_t$.

To model the sequential patterns within user behavior sequences, we incorporate learnable position embeddings $\mathbf{P}_s \in \mathbb{R}^{L_s \times d}$ and $\mathbf{P}_r \in \mathbb{R}^{L_r \times d}$ for the \srcandrec histories, respectively. 
The position-aware history representations are computed as:
$\widehat{\mathbf{E}}_s = \mathbf{E}_{s} + \mathbf{P}_s,
\widehat{\mathbf{E}}_r = \mathbf{E}_{r} + \mathbf{P}_r.$

To further capture the contextual dependencies for the user’s \srcandrec histories, we apply two separate Transformer encoders~\cite{vaswani2017attention}. Each encoder comprises a Multi-Head Self-Attention (MSA) layer followed by a Feed-Forward Network (FFN).
Formally, the encoding procedure is defined as:
\begin{equation}
\label{eq:his_encoder}
\begin{aligned}
\mathbf{H}_s &= \mathrm{FFN}_s(\mathrm{MSA}_s(\mathbf{Q}=\widehat{\mathbf{E}}_s,\mathbf{K}=\widehat{\mathbf{E}}_s,\mathbf{V}=\widehat{\mathbf{E}}_s) + \widehat{\mathbf{E}}_s), \\
\mathbf{H}_r &= \mathrm{FFN}_r(\mathrm{MSA}_r(\mathbf{Q}=\widehat{\mathbf{E}}_r,\mathbf{K}=\widehat{\mathbf{E}}_r,\mathbf{V}=\widehat{\mathbf{E}}_r) + \widehat{\mathbf{E}}_r),
\end{aligned}
\end{equation}
where $\mathbf{H}_s \in \mathbb{R}^{L_s \times d}$ and $\mathbf{H}_r \in \mathbb{R}^{L_r \times d}$ represent the contextualized embeddings of the \srcandrec histories, respectively.
Here, the MSA operates over queries ($\mathbf{Q}$), keys ($\mathbf{K}$), and values ($\mathbf{V}$), all set to the same input sequence.
We apply a causal mask to restrict each position to attend only to its past, preserving temporal order.

\subsubsection{Latent Cross Reasoning}
We propose a latent cross reasoning mechanism that iteratively identifies informative signals between \srcandrec histories. By incorporating cross-attention at each reasoning step, the model progressively refines user interests while explicitly capturing dependencies between the two behavior types.

We initialize the reasoning process with the last hidden states of the encoded histories:
$\mathbf{h}_s^{(0)} = \mathbf{H}_s[-1], ~~ \mathbf{h}_r^{(0)} = \mathbf{H}_r[-1]$,
which serve as the starting point for iterative reasoning.

At each reasoning step $k \in [1, K]$, where $K$ is the total number of reasoning steps, 
the inputs are formed by adding step-specific position embeddings to the previous outputs:
\begin{equation}
\label{eq:reason_step_input}
\hat{\mathbf{e}}_s^{(k)}=\mathbf{h}_s^{(k-1)} + \widetilde{\mathbf{P}}_s[k] \in \mathbb{R}^d,~~
\hat{\mathbf{e}}_r^{(k)}=\mathbf{h}_r^{(k-1)} + \widetilde{\mathbf{P}}_r[k] \in \mathbb{R}^d,
\end{equation}
where 
$\mathbf{h}_s^{(k-1)}, \mathbf{h}_r^{(k-1)} \in \mathbb{R}^d$ are the outputs from previous step.
$\widetilde{\mathbf{P}}_s, \widetilde{\mathbf{P}}_r \in \mathbb{R}^{K \times d}$
are learnable position embeddings that differentiate reasoning inputs from historical behaviors.

The search reasoning proceeds in two stages. First, a self-attention layer summarizes the search history and previous reasoning inputs:
\begin{equation}
\label{eq:src_reason_msa}
\mathbf{f}_s^{(k)} = \mathrm{MSA}_s(\mathbf{Q}=\hat{\mathbf{e}}_s^{(k)},\mathbf{K}=\widehat{\mathbf{E}}_s^{(k)},\mathbf{V}=\widehat{\mathbf{E}}_s^{(k)}) + \hat{\mathbf{e}}_s^{(k)},
\end{equation}
where $\mathbf{f}_s^{(k)} \in \mathbb{R}^d$. $\widehat{\mathbf{E}}_s^{(k)}$ includes both the original search history and past reasoning inputs. This allows the model to refine the current input based on intra-history context.

Then, a cross-attention layer integrates signals from the recommendation side to update the search representation:
\begin{equation}
\label{eq:src_reason_mca}
\mathbf{h}_s^{(k)}= \mathrm{FFN}_s(\mathrm{MCA}_s(\mathbf{Q}=\mathbf{f}_s^{(k)},\mathbf{K}=\widehat{\mathbf{E}}_r^{(k)},\mathbf{V}=\widehat{\mathbf{E}}_r^{(k)}) + \mathbf{f}_s^{(k)}),
\end{equation}
where the Multi-Head Cross-Attention (MCA) enables the model to selectively extract recommendation-side information relevant to the current search reasoning step.
Please note that $\mathrm{MSA}_s$ and $\mathrm{FFN}_s$ share the same parameters as defined in Eq.~\eqref{eq:his_encoder}.
The recommendation reasoning is symmetric:
\begin{equation}
\label{eq:rec_reason}
\begin{aligned}
\mathbf{f}_r^{(k)} &= \mathrm{MSA}_r(\mathbf{Q}=\hat{\mathbf{e}}_r^{(k)},\mathbf{K}=\widehat{\mathbf{E}}_r^{(k)},\mathbf{V}=\widehat{\mathbf{E}}_r^{(k)}) + \hat{\mathbf{e}}_r^{(k)},\\
\mathbf{h}_s^{(k)}&= \mathrm{FFN}_r(\mathrm{MCA}_r(\mathbf{Q}=\mathbf{f}_r^{(k)},\mathbf{K}=\widehat{\mathbf{E}}_s^{(k)},\mathbf{V}=\widehat{\mathbf{E}}_s^{(k)}) + \mathbf{f}_r^{(k)}),
\end{aligned}
\end{equation}
allowing the recommendation encoder to incorporate information from the updated search history.

Here, the attention contexts $\widehat{\mathbf{E}}_s^{(k)} \in \mathbb{R}^{(L_s+k)\times d}$ and $\widehat{\mathbf{E}}_r^{(k)} \in \mathbb{R}^{(L_r+k)\times d}$ are updated at each step by concatenating the original histories with all previous reasoning inputs:
\begin{equation}
\label{eq:reason_input}
\begin{aligned}
\widehat{\mathbf{E}}_s^{(k)} = [\widehat{\mathbf{E}}_s^{\mathsf{T}}, \hat{\mathbf{e}}_s^{(1)},\ldots,\hat{\mathbf{e}}_s^{(k)}]^{\mathsf{T}} , 
\widehat{\mathbf{E}}_r^{(k)} = [\widehat{\mathbf{E}}_r^{\mathsf{T}}, \hat{\mathbf{e}}_r^{(1)},\ldots,\hat{\mathbf{e}}_r^{(k)}]^{\mathsf{T}} .
\end{aligned}
\end{equation}
This design allows the reasoning module to access both the static behavior sequences and dynamically generated reasoning inputs.
After $K$ steps, we aggregate all representations to form the final history embeddings:
\begin{equation}
\label{eq:reason_output}
\begin{aligned}
\mathbf{H}_s^{(K)}&= [\mathbf{H}_s^{\mathsf{T}}, \mathbf{h}_s^{(1)},\ldots,\mathbf{h}_s^{(K)}]^{\mathsf{T}} \in \mathbb{R}^{(L_s+K)\times d}, \\
\mathbf{H}_r^{(K)}&= [\mathbf{H}_r^{\mathsf{T}}, \mathbf{h}_r^{(1)},\ldots,\mathbf{h}_r^{(K)}]^{\mathsf{T}} \in \mathbb{R}^{(L_r+K)\times d}.
\end{aligned}
\end{equation}
These enriched representations integrate original behavior and step-wise cross-history reasoning, enabling deeper modeling of user interest.

\subsubsection{Target-Aware Reasoning Aggregation}
Following latent cross reasoning, we compute the similarity between the candidate item embedding and all hidden states from both histories and reasoning steps, then adaptively aggregate them. This enables the model to focus on the most informative reasoning step, which may vary across samples. Formally:
\begin{equation}
\label{eq:target_agg}
\begin{aligned}
\mathbf{w}_s &= \mathrm{Att}(\mathbf{Q}=\mathbf{e}_{v_{L+1}}, \mathbf{K}=\mathbf{H}_s^{(K)}, \mathbf{V}=\mathbf{H}_s^{(K)}) \in \mathbb{R}^d,\\
\mathbf{w}_r &= \mathrm{Att}(\mathbf{Q}=\mathbf{e}_{v_{L+1}}, \mathbf{K}=\mathbf{H}_r^{(K)}, \mathbf{V}=\mathbf{H}_r^{(K)}) \in \mathbb{R}^d,
\end{aligned}
\end{equation}
where $\mathrm{Att}(\cdot)$ denotes the attention function used to compute similarity.
The embedding $\mathbf{e}_{v_{L+1}}$ of the candidate item serves as the query, while 
$\mathbf{H}_s^{K}$ and $\mathbf{H}_r^{K}$ 
are used as the key and value. 
The outputs $\mathbf{w}_s$ and $\mathbf{w}_r$ provide target-aware user representations for final prediction.

\subsection{Supervised Pre-training}
After encoding and reasoning over the \srcandrec histories, we conduct supervised pre-training to equip the model with reasoning and recommendation capabilities. A contrastive loss encourages the hidden states after cross reasoning to be closer to the target, and a recommendation loss aligns the model with the recommendation task.

\subsubsection{Target-Aware Contrastive Learning}
To improve cross-attention’s ability to extract useful information from the complementary behavior sequence, we introduce a target-aware contrastive learning (\textbf{TCL}) objective. It encourages representations obtained with cross reasoning to be closer to the target item than those without, guiding the model to capture complementary and prediction-relevant signals.

Formally, let $\mathbf{H}_s^{(K)}$, $\mathbf{H}_r^{(K)}$ be the final representations after $K$ cross reasoning steps (Eq.~\eqref{eq:reason_output}), and $\widetilde{\mathbf{H}}_s^{(K)}$, $\widetilde{\mathbf{H}}_r^{(K)}$ the counterparts without cross-attention (by removing cross-attention modules in Eq.~\eqref{eq:src_reason_mca}, Eq.~\eqref{eq:rec_reason}).
We use the final-step hidden states $\mathbf{h}_s^{(K)}, \mathbf{h}_r^{(K)}, \tilde{\mathbf{h}}_s^{(K)}, \tilde{\mathbf{h}}_r^{(K)} \in \mathbb{R}^d$ for contrastive learning, as they aggregate the information from all previous reasoning steps.
The loss is defined as:
\begin{equation}
\label{eq:target_cl}
\begin{cases}
\mathcal{L}_{\mathrm{TCL\text{-}S}} = \mathrm{max}\{d(\mathbf{e}_{v_{L+1}},\mathbf{h}_s^{(K)}) - d(\mathbf{e}_{v_{L+1}},\tilde{\mathbf{h}}_s^{(K)}) + m,0\}, \\
\mathcal{L}_{\mathrm{TCL\text{-}R}} = \mathrm{max}\{d(\mathbf{e}_{v_{L+1}},\mathbf{h}_r^{(K)}) - d(\mathbf{e}_{v_{L+1}},\tilde{\mathbf{h}}_r^{(K)}) + m,0\}, \\
\mathcal{L}_{\mathrm{TCL}} = \mathcal{L}_{\mathrm{TCL\text{-}S}} + \mathcal{L}_{\mathrm{TCL\text{-}R}}.
\end{cases}
\end{equation}
where $d(\cdot,\cdot)$ is a distance metric (e.g., cosine or Euclidean), and $m > 0$ is a margin. This objective explicitly promotes the extraction of complementary information relevant to the target, enhancing the effectiveness of cross reasoning.

\subsubsection{Prediction and Training for Recommendation}
Finally, we concatenate the user embedding $\mathbf{e}_u$, the aggregated historical representations $\mathbf{w}_s$ and $\mathbf{w}_r$ derived from Eq.~\eqref{eq:target_agg},
and the candidate item embedding $\mathbf{e}_{v_{L+1}}$. 
The resulting vector is passed through a multi-layer perceptron (MLP) to compute the predicted preference score:
\begin{equation}
\label{eq:predict}
    \hat{y}_{u,v_{L+1}} = \mathrm{MLP}(\mathrm{CONCAT}(\mathbf{e}_u,\mathbf{w}_s,\mathbf{w}_r,\mathbf{e}_{v_{L+1}})),
\end{equation}
where $\hat{y}_{u,v_{L+1}}$ denotes the predicted preference score of user $u$ on item $v_{L+1}$. $\mathrm{CONCAT}(\cdot)$ represents vector concatenation.
Following prior works~\cite{UniSAR}, we optimize the recommendation model using binary cross-entropy loss:
\begin{equation}
\label{eq:click_loss}
\begin{aligned}
\mathcal{L}_{\mathrm{rec}} = -\frac{1}{|\mathcal{D}|} \sum_{(u,v_{L+1}) \in \mathcal{D}} \left[ y_{u,v_{L+1}}\mathrm{log}(\hat{y}_{u,v_{L+1}}) + \right. \\
\left. (1-y_{u,v_{L+1}})\mathrm{log}(1-\hat{y}_{u,v_{L+1}}) \right],  
\end{aligned}
\end{equation}
where $\mathcal{D}$ is the training set of user-item pairs, and $y_{u,v_{L+1}} \in \{0,1\}$ denotes the ground-truth label.

The overall training objective combines the recommendation loss with the target-aware contrastive loss (Eq.~\eqref{eq:target_cl}) and $L_2$ regularization on model parameters:
\begin{equation}
\label{eq:total_loss}
\begin{aligned}
\mathcal{L}_{\mathrm{Total}} = \mathcal{L}_{\mathrm{rec}} + \lambda_{\mathrm{TCL}} \mathcal{L}_{\mathrm{TCL}} + \lambda_{\mathrm{Reg}}||\Phi||^2,
\end{aligned}
\end{equation}
where $\lambda_{\mathrm{TCL}}$ and $\lambda_{\mathrm{Reg}}$ are hyper-parameters that weight the contrastive loss and regularization term, respectively, and $\Phi$ denotes all trainable parameters.

\subsection{Reinforcement Learning}
While supervised pre-training provides basic reasoning abilities, its deterministic optimization limits exploration. To overcome this, we incorporate reinforcement learning with Group Relative Policy Optimization (GRPO)~\cite{shao2024deepseekmath,guo2025deepseek}, allowing the model to explore diverse reasoning trajectories and optimize them for improved recommendation performance.

\subsubsection{Trajectory Rollout}
To encourage exploration, we inject Gaussian noise into the initial hidden states, allowing the model to generate diverse reasoning trajectories. Specifically, we sample $N$ trajectories $\{\mathcal{T}_i\}_{i=1}^{N}$ from the old policy $\pi_{\mathrm{old}}$ (i.e., the policy before update), where each trajectory is defined as:
\begin{equation}
\label{eq:trajectory}
\mathcal{T}_i = \left[ (\mathbf{h}_{s,i}^{(0)}, \mathbf{h}_{r,i}^{(0)}), \ldots, (\mathbf{h}_{s,i}^{(K)}, \mathbf{h}_{r,i}^{(K)}) \right].
\end{equation}
where $(\mathbf{h}_{s,i}^{(k)}, \mathbf{h}_{r,i}^{(k)})$ denote 
the reasoning states at step $k$ for trajectory $i$.
To ensure exploration, we perturb the initial states as:
\begin{equation}
\label{eq:reason_noise}
\mathbf{h}_{s,i}^{(0)} = \mathbf{h}_s^{(0)} + \mathbb{I}_{[i > 1]} \cdot \gamma \cdot \boldsymbol{\epsilon}_{s,i}, ~~
\mathbf{h}_{r,i}^{(0)} = \mathbf{h}_r^{(0)} + \mathbb{I}_{[i > 1]} \cdot \gamma \cdot \boldsymbol{\epsilon}_{r,i},
\end{equation}
where $\mathbb{I}_{[i > 1]}$ indicates whether noise is added, ensuring the first trajectory remains deterministic. $\boldsymbol{\epsilon}_{s,i}, \boldsymbol{\epsilon}_{r,i} \sim \mathcal{N}(\mathbf{0}, \sigma^2 \mathbf{I})$ are Gaussian noise vectors, and $\gamma$ controls their magnitude.
This design enables exploration by perturbing reasoning while preserving the original trajectory for stability.

\subsubsection{Reward Design}
To better align the model’s optimization with the recommendation objective, we adopt standard ranking metrics (e.g., NDCG, Hit Ratio) as reward signals in reinforcement learning.
For each sampled reasoning trajectory $\mathcal{T}_i$, we compute the relevance scores of all candidate items using Eq.~\eqref{eq:target_agg} and Eq.~\eqref{eq:predict}, and obtain its corresponding ranking list $L_i$. The reward is then defined as:
$r_i = \mathrm{Eval}(L_i)$, where $\mathrm{Eval}(\cdot)$ denotes the ranking metric used to evaluate the quality of $L_i$.

\subsubsection{Policy Optimization}
We adopt the GRPO algorithm~\cite{shao2024deepseekmath} to update the current policy $\pi_{\Phi}$ (i.e., the model being trained) by maximizing expected rewards while regularizing the policy update with a KL divergence constraint relative to a reference policy $\pi_{\mathrm{ref}}$ (i.e., the initially pre-trained model). The GRPO objective is formulated as:
\begin{equation}
\label{eq:grpo}
\begin{aligned}
\mathcal{J}_{\mathrm{GRPO}}(\Phi) = 
\mathbb{E}_{\{\mathcal{T}_i\}_{i=1}^{N} \sim \pi_{\mathrm{old}}} 
 \left[ \frac{1}{N} \sum_{i=1}^{N} \mathcal{J}_i \right], \\
\mathcal{J}_{i} = \frac{\pi_{\Phi}(v_{L+1} \mid \mathcal{T}_i)}{\pi_{\mathrm{old}}(v_{L+1} \mid \mathcal{T}_i)} A_i - \lambda_{\mathrm{KL}} \mathbb{D}_{\mathrm{KL}}(\pi_\Phi || \pi_{\mathrm{ref}}),
\end{aligned}
\end{equation}
where $\lambda_{\mathrm{KL}}$ controls the influence of the KL term, and the KL divergence is computed as follows:
\begin{equation}
\label{eq:kl_div}
\mathbb{D}_{\mathrm{KL}}(\pi_\Phi || \pi_{\mathrm{ref}}) = 
\frac{\pi_{\mathrm{ref}}(v_{L+1} \mid \mathcal{T}_i)}{\pi_{\Phi}(v_{L+1} \mid \mathcal{T}_i)} - 
\mathrm{log} \frac{\pi_{\mathrm{ref}}(v_{L+1} \mid \mathcal{T}_i)}{\pi_{\Phi}(v_{L+1} \mid \mathcal{T}_i)} - 1,
\end{equation}
where $v_{L+1}$ denotes the target item, and $\pi(v_{L+1} \mid \mathcal{T}_i)$ is the probability of predicting the target item given the reasoning trajectory. This divergence quantifies the deviation from the reference policy and constrains excessive policy drift.

The advantage term $A_i$ in Eq.~\eqref{eq:grpo} is computed as the standardized reward:
\begin{equation}
\label{eq:advantage}
A_i = \frac{r_i - \mathrm{mean}(\{r_1,r_2,\ldots,r_N\})}{\mathrm{std}(\{r_1,r_2,\ldots,r_N\})},
\end{equation}
which reflects the relative improvement of trajectory $\mathcal{T}_i$ compared to other sampled trajectories, thereby guiding the policy update toward more effective reasoning paths.

\begin{table}[t]
\small
\centering
\caption{Statistics of the datasets used in this paper. ``S'' and ``R'' denote search and recommendation, respectively.}
 \resizebox{1.0\columnwidth}{!}{
\begin{tabular}{cccccc}
\toprule
Dataset & \#Users & \#Items & \#Queries & \#Interaction-S &\#Interaction-R  \\
\midrule
Qilin &15,482 &1,983,938 &44,820 &969,866 &1,438,435 \\
KuaiSAR-Small & 25,877 & 4,195,529 & 267,608 & 3,171,231 & 7,493,101\\
KuaiSAR-Large & 25,877 & 6,890,707 & 453,667 & 5,059,169 & 14,605,716 \\
\bottomrule
\end{tabular}
}
\label{tab:dataStatistics}   
\end{table}

\section{Experiments}
We conducted experiments to validate the effectiveness of \ourname.

\begin{table*}[t]
\centering
\caption{
Overall recommendation performance of different methods across all datasets.
H@$k$ and N@$k$ refer to HR@$k$ and NDCG@$k$, respectively.
The best and second-best results are marked in bold and underlined fonts, respectively.
``*'' indicates that the improvement over the second-best method is statistically significant ($t$-test, $p < 0.05$).
}
\resizebox{1.0\linewidth}{!}{
\begin{tabular}{
lc
ccccc
ccccc
ccccc
}
\toprule
\multicolumn{1}{c}{\multirow{2}{*}{Category}} & 
\multicolumn{1}{c}{\multirow{2}{*}{Methods}} & 
\multicolumn{5}{c}{Qilin} & 
\multicolumn{5}{c}{KuaiSAR-Small} & 
\multicolumn{5}{c}{KuaiSAR-Large} \\
\cmidrule(l){3-7} \cmidrule(l){8-12} \cmidrule(l){13-17} 
\multicolumn{1}{c}{}  & \multicolumn{1}{c}{}  
&H@1 &H@5 &H@10 &N@5 &N@10
&H@1 &H@5 &H@10 &N@5 &N@10
&H@1 &H@5 &H@10 &N@5 &N@10 \\
\midrule
\multirow{7}*{Recommendation} 
&GRU4Rec &0.1262 &0.3397 &0.4796 &0.2360 &0.2811 &0.0469 &0.1781 &0.2947 &0.1123 &0.1498 &0.0950 &0.3299 &0.5199 &0.2128 &0.2739 \\
&SASRec &0.1318 &0.3473 &0.4871 &0.2423 &0.2873 &0.0501 &0.1820 &0.2872 &0.1162 &0.1500 &0.1035 &0.3583 &0.5599 &0.2310 &0.2959 \\
&BERT4Rec &0.1325 &0.3531 &0.4877 &0.2475 &0.2909 &0.0489 &0.1801 &0.2841 &0.1145 &0.1480 &0.1075 &0.3613 &0.5594 &0.2349 &0.2987 \\
&CL4SRec &0.1412 &0.3574 &0.4879 &0.2527 &0.2946 &0.0514 &0.1833 &0.2875 &0.1175 &0.1510 &0.1080 &0.3685 &0.5690 &0.2390 &0.3036 \\
&ReaRec (ERL) &0.1393 &0.3562 &0.4956 &0.2502 &0.2951 &0.0509 &0.1843 &0.2930 &0.1176 &0.1526 &0.1140 &0.3812 &0.5747 &0.2486 &0.3110 \\
&ReaRec (PRL) &0.1451 &0.3647 &\underline{0.5032} &0.2579 &0.3025 &0.0603 &0.2207 &0.3637 &0.1404 &0.1863 &0.1078 &0.3730 &0.5732 &0.2409 &0.3054 \\
&LARES &0.1426 &0.3596 &0.5031 &0.2541 &0.3003 &0.0514 &0.1860 &0.3050 &0.1186 &0.1567 &0.1086 &0.3702 &0.5553 &0.2405 &0.3002 \\
\hline
\multirow{8}*{\makecell[l]{Search Enhanced \\ Recommendation}}
&NRHUB &0.1313 &0.3464 &0.4871 &0.2423 &0.2876 &0.0656 &0.2311 &0.3642 &0.1487 &0.1914 &0.1036 &0.3469 &0.5209 &0.2264 &0.2825\\
&Query-SeqRec &0.1394 &0.3495 &0.4855 &0.2470 &0.2907 &0.0511 &0.1869 &0.2940 &0.1192 &0.1536 &0.1084 &0.3635 &0.5642 &0.2366 &0.3013 \\
&JSR &0.1435 &0.3622 &0.4931 &0.2553 &0.2975 &0.0520 &0.1886 &0.3007 &0.1205 &0.1564 &0.1241 &0.4001 &0.5941 &0.2633 &0.3258 \\
&USER &0.1452 &0.3629 &0.4888 &0.2573 &0.2981 &0.0659 &0.2285 &0.3553 &0.1475 &0.1883 &0.1340 &0.3825 &0.5382 &0.2607 &0.3109 \\
&SESRec &0.1431 &0.3558 &0.4892 &0.2536 &0.2966 &0.0804 &0.2610 &0.3942 &0.1714 &0.2142 &0.1556 &0.4395 &0.6062 &0.3005 &0.3544\\
&UnifiedSSR &0.1373 &0.3604 &0.4921 &0.2521 &0.2946 &0.0471 &0.1773 &0.2898 &0.1123 &0.1484 &0.1098 &0.3661 &0.5510 &0.2392 &0.2988\\
&UniSAR  &\underline{0.1528} &\underline{0.3689} &{0.4963} &\underline{0.2643} &\underline{0.3053} &\underline{0.0874} &\underline{0.2749} &\underline{0.4104} &\underline{0.1821} &\underline{0.2257} &\underline{0.1676} &\underline{0.4597} &\underline{0.6274} &\underline{0.3174} &\underline{0.3716}\\
&\textbf{\ourname}  &\textbf{0.1675}* &\textbf{0.3936}* &\textbf{0.5114}* &\textbf{0.2844}* &\textbf{0.3224}* &\textbf{0.0968}* &\textbf{0.3253}* &\textbf{0.4894}* &\textbf{0.2120}* &\textbf{0.2648}* &\textbf{0.1810}* &\textbf{0.4902}* &\textbf{0.6553}* &\textbf{0.3396}* &\textbf{0.3930}*\\
\bottomrule
\end{tabular} 
}
\label{tab:rec_result}
\end{table*}

\subsection{Experimental Setup}

\subsubsection{Dataset}
We conducted experiments on the following datasets, with their statistics summarized in Table~\ref{tab:dataStatistics}.
(1)~\textbf{Qilin}~\cite{chen2025qilin}: A dataset from Xiaohongshu,
containing user interactions in both \srcandrec scenarios. We follow~\cite{BERT4REC,SESRec} and apply the leave-one-out strategy for splitting.
(2)~\textbf{KuaiSAR}~\cite{sun2023kuaisar}: A dataset from Kuaishou,
covering cross-scenario interactions. We use both the \textbf{small} and \textbf{large} versions, and split the data chronologically following~\cite{UniSAR}.
Details of data processing and splitting are provided in the supplementary~material.

\subsubsection{Baselines}
We compare our method with the following two categories of baselines:
(1)~\emph{Recommendation}:
\textbf{GRU4Rec}~\cite{GRU4REC};
\textbf{SASRec}~\cite{SASREC};
\textbf{BERT4Rec}~\cite{BERT4REC};
\textbf{CL4SRec}~\cite{xie2022contrastive};
\textbf{ReaRec (ERL)}~\cite{tang2025think}; 
\textbf{ReaRec (PRL)}~\cite{tang2025think};
\textbf{LARES}~\cite{liu2025lares};
(2)~\emph{Search enhanced recommendation}:
\textbf{NRHUB}~\cite{NRHUB};
\textbf{Query-SeqRec}~\cite{Query_SeqRec};
\textbf{JSR}~\cite{JSR};
\textbf{USER}~\cite{USER};
\textbf{SESRec}~\cite{SESRec};
\textbf{UnifiedSSR}~\cite{xie2024unifiedssr};
\textbf{UniSAR}~\cite{UniSAR}.
Further details are in the supplementary material.

\subsubsection{Evaluation \& Implementation}
We evaluate models using Hit Ratio (HR)@$\{1,5,10\}$ and Normalized Discounted Cumulative Gain (NDCG)@$\{5,10\}$, following standard protocols~\cite{UniSAR}. Each test item is ranked against 99 randomly sampled negatives~\cite{UniSAR}.
We use HR@1 as the reward for reinforcement learning.
For more details on evaluation and model implementation, please refer to the supplementary material.

\subsection{Overall Performance}
Table~\ref{tab:rec_result} presents the recommendation performance on the three datasets. We observe that:

\begin{itemize}[leftmargin=1em]
    \item First, we observe that \ourname achieves state-of-the-art performance compared to existing recommendation models and search-enhanced recommendation methods. This demonstrates the effectiveness of latent cross reasoning in extracting useful signals from search behaviors to enhance recommendation, and vice versa.
    \item Second, we observe that search-enhanced recommendation models, such as \ourname and UniSAR, significantly outperform traditional recommendation models, demonstrating the effectiveness of leveraging search data. However, some models like NRHUB perform worse than traditional baselines, indicating that search signals cannot be incorporated in a naive or straightforward manner.
    \item Finally, we observe that models based on latent reasoning, including ReaRec, LARES, and \ourname, generally outperform comparable recommendation or search-enhanced recommendation models without latent reasoning. This highlights the effectiveness of incorporating latent reasoning, which enables the model to progressively better capture and understand user interests.
\end{itemize}

\begin{table}[!t]
\small
\centering
\caption{
Ablation study of \ourname on Qilin and KuaiSAR-Small, where each module is incrementally added to the Base model.
}
\renewcommand{\arraystretch}{1.2}
\resizebox{1.0\columnwidth}{!}{
\begin{tabular}
{ll
cccc}
\toprule
\multicolumn{2}{c}{Variants} & 
\multicolumn{2}{c}{Qilin} & 
\multicolumn{2}{c}{KuaiSAR-Small} \\ 
\cmidrule(l){1-2}
\cmidrule(l){3-4} \cmidrule(l){5-6}
\# &Model
&H@5 &N@5
&H@5 &N@5  \\ 
\midrule
\ding{172} &Base &0.3350 &0.2294 &0.2551 &0.1661 \\
\hdashline
\ding{173} &\ding{172}+LCR &0.3421 &0.2346 &0.2643 &0.1712 \\
\hdashline
\ding{174} &\ding{173}+TRA &0.3804 &0.2702 &0.2954 &0.1949 \\
\hdashline
\ding{175} &\ding{174}+$\mathcal{L}_{\mathrm{TCL}}$ &0.3793 &0.2738 &0.3185 &0.2107 \\
\hdashline
\ding{176} &\textbf{\ourname} (\ding{175}+RL ) &\textbf{0.3936} &\textbf{0.2844} &\textbf{0.3253} &\textbf{0.2120} \\
\bottomrule
\end{tabular}
} 
\label{tab:ablation_result}
\end{table}

\begin{table}[!t]
\small
\centering
\caption{
Inference time across different reasoning steps on the Qilin test set.
``Increase (\%)'' denotes the relative time increase compared to ``Step-0''.
Measured on a single A6000 48GB GPU.
Time is measured in seconds (s).
}
\renewcommand{\arraystretch}{1.2}
\resizebox{1.0\columnwidth}{!}{
\begin{tabular}
{l
cccccc}
\toprule
&Step-0 &Step-1 &Step-2 &Step-3 &Step-4 &Step-5 \\
\midrule
Time (s) &4.4581 &4.6949 &5.0641 &5.5271 &5.7207 &6.2515  \\
Increase (\%) &\textemdash &5.31 &13.59 &23.98 &28.32 &40.23 \\
\bottomrule
\end{tabular}
} 
\label{tab:latency}
\end{table}

\subsection{Ablation Study}
We conduct ablation studies on Qilin and KuaiSAR-Small to evaluate the contribution of each module in \ourname. As shown in Table~\ref{tab:ablation_result}, modules are added incrementally.
``Base'' uses a Transformer over \srcandrec histories, with mean pooling on the final hidden states for~prediction.

\subsubsection{Latent Cross Reasoning (LCR)}
Introducing LCR yields consistent performance improvements, indicating that it effectively guides the cross-attention mechanism to extract informative signals from search behaviors to enhance recommendation representations, and vice versa. This shows that LCR facilitates the mutual enhancement between \srcandrec by allowing one to leverage signals from the other.

\subsubsection{Target-Aware Reasoning Aggregation (TRA)}
Adding TRA results in a significant performance boost, highlighting the importance of adaptively aggregating reasoning states with respect to the target. Since different samples may benefit from different reasoning depths, TRA enables the model to automatically select the most appropriate reasoning step, avoiding manual tuning and improving prediction accuracy for each sample.

\subsubsection{Target-Aware Contrastive Learning ($\mathcal{L}_{\mathrm{TCL}}$)}
The incorporation of $\mathcal{L}_{\mathrm{TCL}}$ further enhances performance. This indicates that contrastive learning helps the model refine cross-attentive representations by encouraging the extraction of predictive signals from search histories to support recommendation interests, and vice versa.

\subsubsection{Reinforcement Learning (RL)}
Finally, incorporating RL brings additional gains by enabling the model to explore diverse reasoning trajectories. By rewarding paths that lead to accurate predictions and penalizing suboptimal ones, the model learns more effective reasoning strategies beyond those captured by supervised pre-training.

\begin{figure}[t]
 \centering
 \subfigure[Histogram of Distance]{
    \label{fig:reason_hidden_histogram}
    \includegraphics[width=0.45\columnwidth]{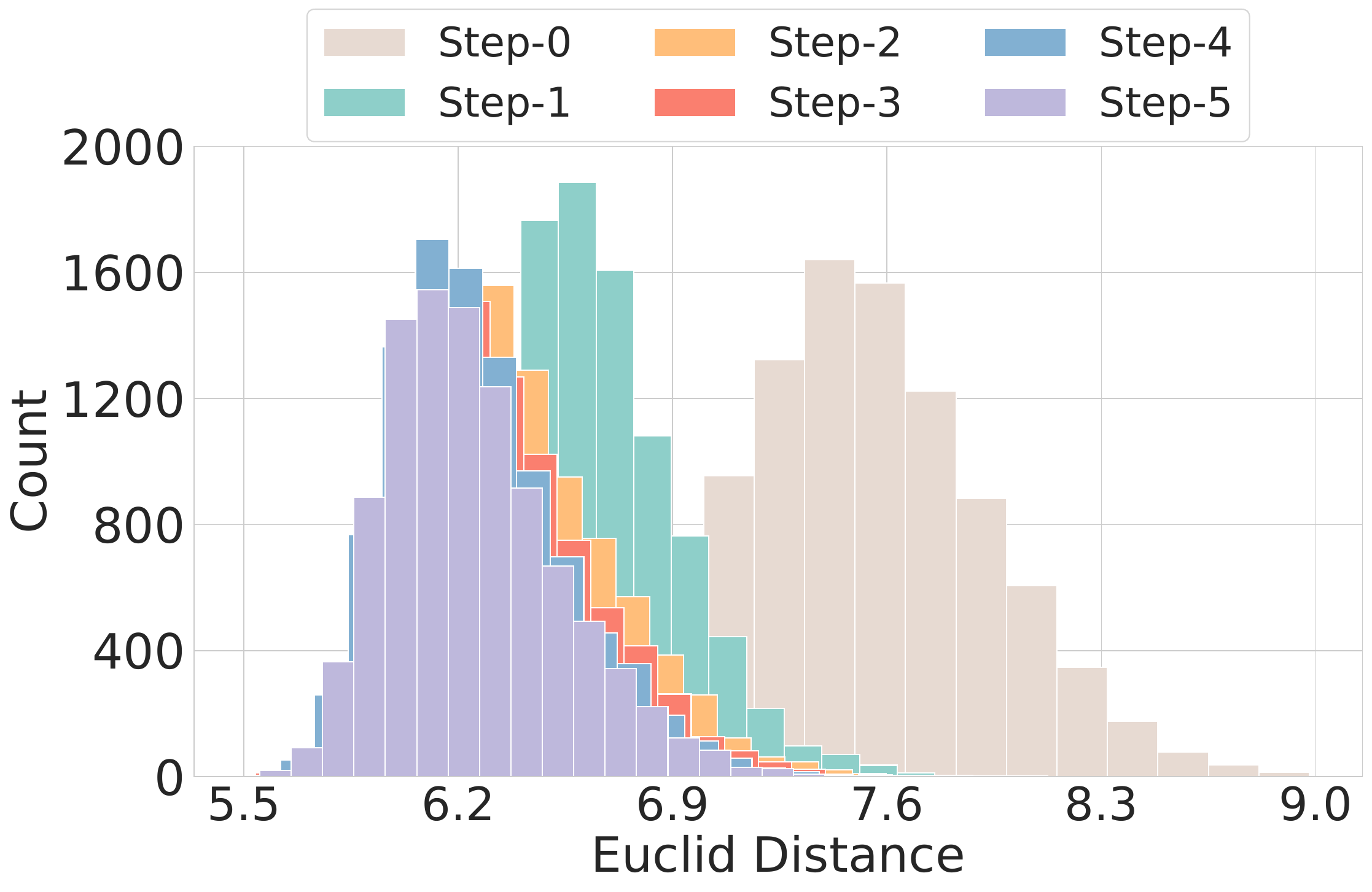}
 }
 \subfigure[t-SNE Visualization]{
    \label{fig:reason_hidden_tsne}
    \includegraphics[width=0.45\columnwidth]{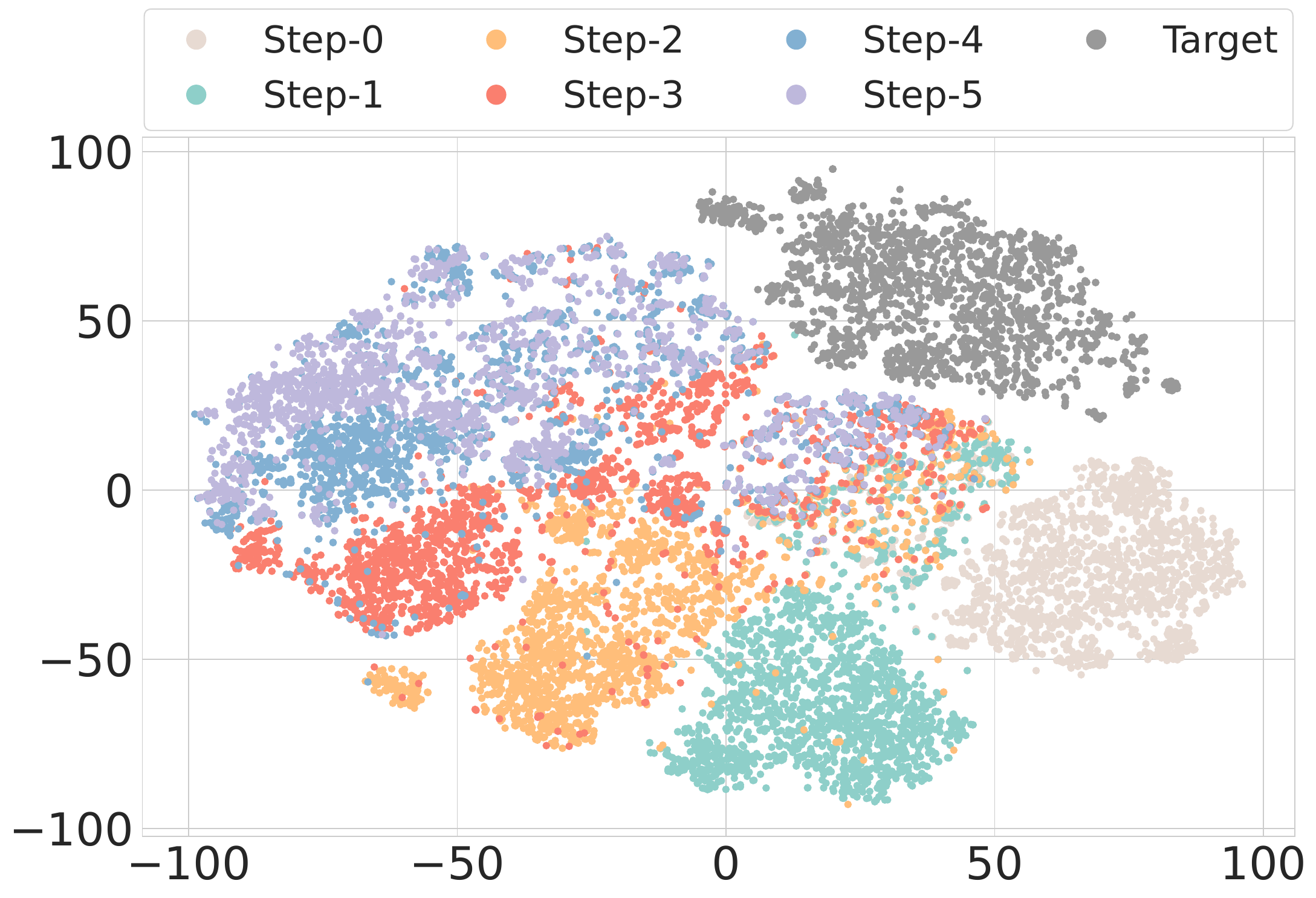}
 }
 \caption{
 Analysis of hidden representations across reasoning steps:
(a)~Histogram of distances between hidden states and target item embeddings at each reasoning step.
(b)~t-SNE visualization of hidden states across reasoning steps and target embeddings.
Reduced distances indicate increased similarity to the target item.
 }
 \label{fig:reason_hidden}
\end{figure}

\subsection{Experimental Analysis}
We further conducted experimental analysis to verify the effectiveness of different modules.

\subsubsection{Effectiveness of Latent Cross Reasoning}
To verify whether reasoning can extract useful information from search history for recommendation,
we analyze the relationship between the hidden representations at different reasoning steps and the target item embeddings, as shown in Figure~\ref{fig:reason_hidden}. We compute the distances between hidden states and target embeddings across reasoning steps and visualize their distributions using t-SNE~\cite{van2008visualizing}. As the number of reasoning steps increases, the hidden representations become progressively closer to the target, while the initial representation without reasoning (Step = 0) shows the largest distance. These results demonstrate that the reasoning process effectively distills useful information from search history to benefit recommendation.

\subsubsection{Analysis of Inference Latency}
We analyze inference latency after introducing reasoning, as shown in Table~\ref{tab:latency}. While latency increases with more reasoning steps, the reasoning mainly processes historical behaviors whose results can be precomputed offline, minimizing online overhead. Overall, the latency increase is acceptable and is outweighed by substantial performance gains, validating the effectiveness of our approach.

\begin{figure}[t]
 \centering
 \subfigure[w/o $\mathcal{L}_{\mathrm{TCL}}$]{
    \label{fig:rec_wo_tcl}
    \includegraphics[width=0.45\columnwidth]{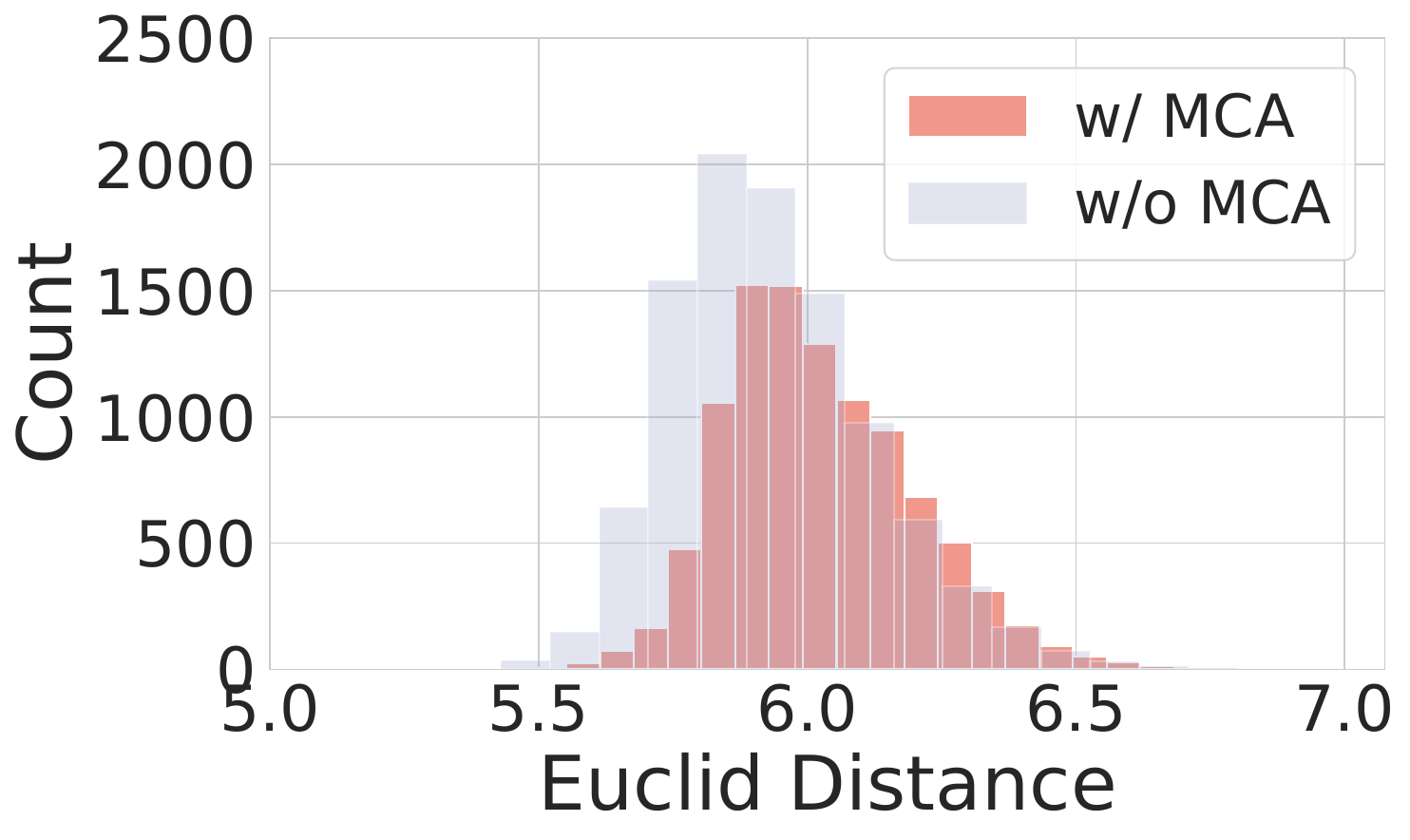}
 }
 \subfigure[w/ $\mathcal{L}_{\mathrm{TCL}}$]{
    \label{fig:rec_w_tcl}
    \includegraphics[width=0.45\columnwidth]{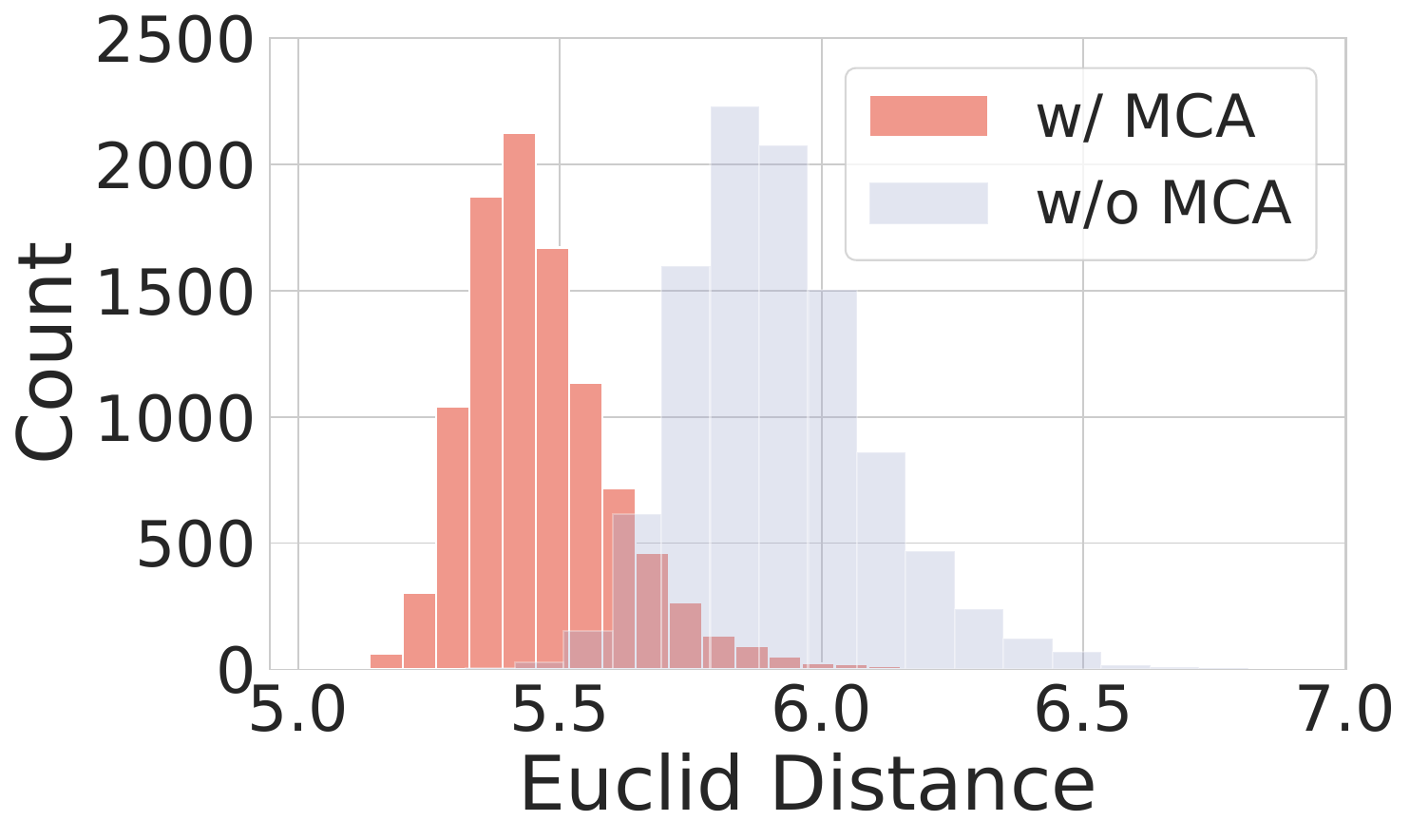}
 }
 \caption{
 Histogram of the distance distribution between the hidden state output of the final reasoning step and the target item,
 comparing results without (``w/o'') and with (``w/'') the $\mathcal{L}_{\mathrm{TCL}}$ loss. 
 Smaller distances indicate greater similarity to the target item.
 }
 \label{fig:tcl_impact}
\end{figure}

\begin{figure}[t]
 \centering
 \subfigure[Qilin]{
    \label{fig:qilin_step}
    \includegraphics[width=0.45\columnwidth]{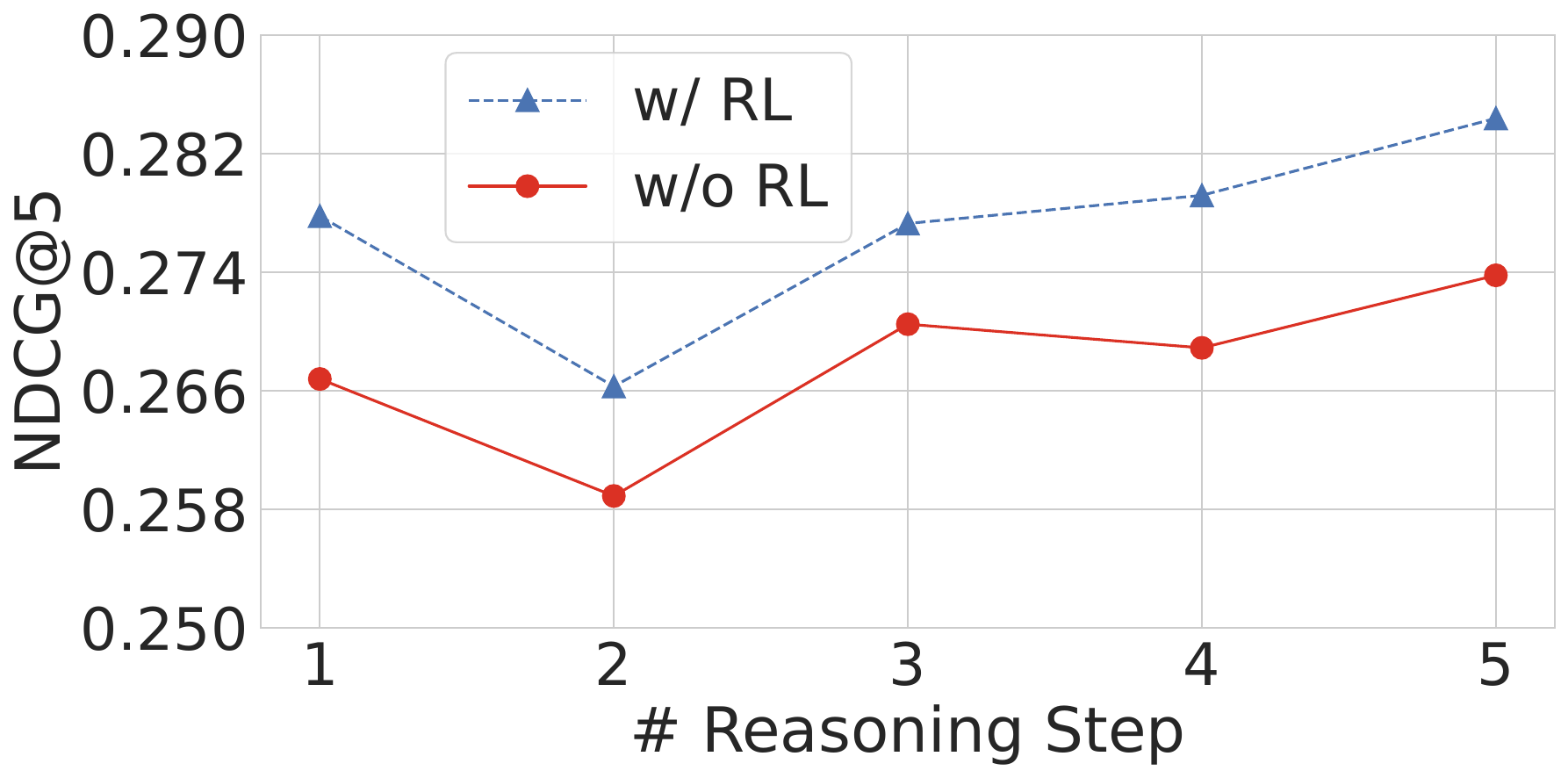}
 }
\subfigure[KuaiSAR-Small]{
    \label{fig:kuaisar_small_step}
    \includegraphics[width=0.45\columnwidth]{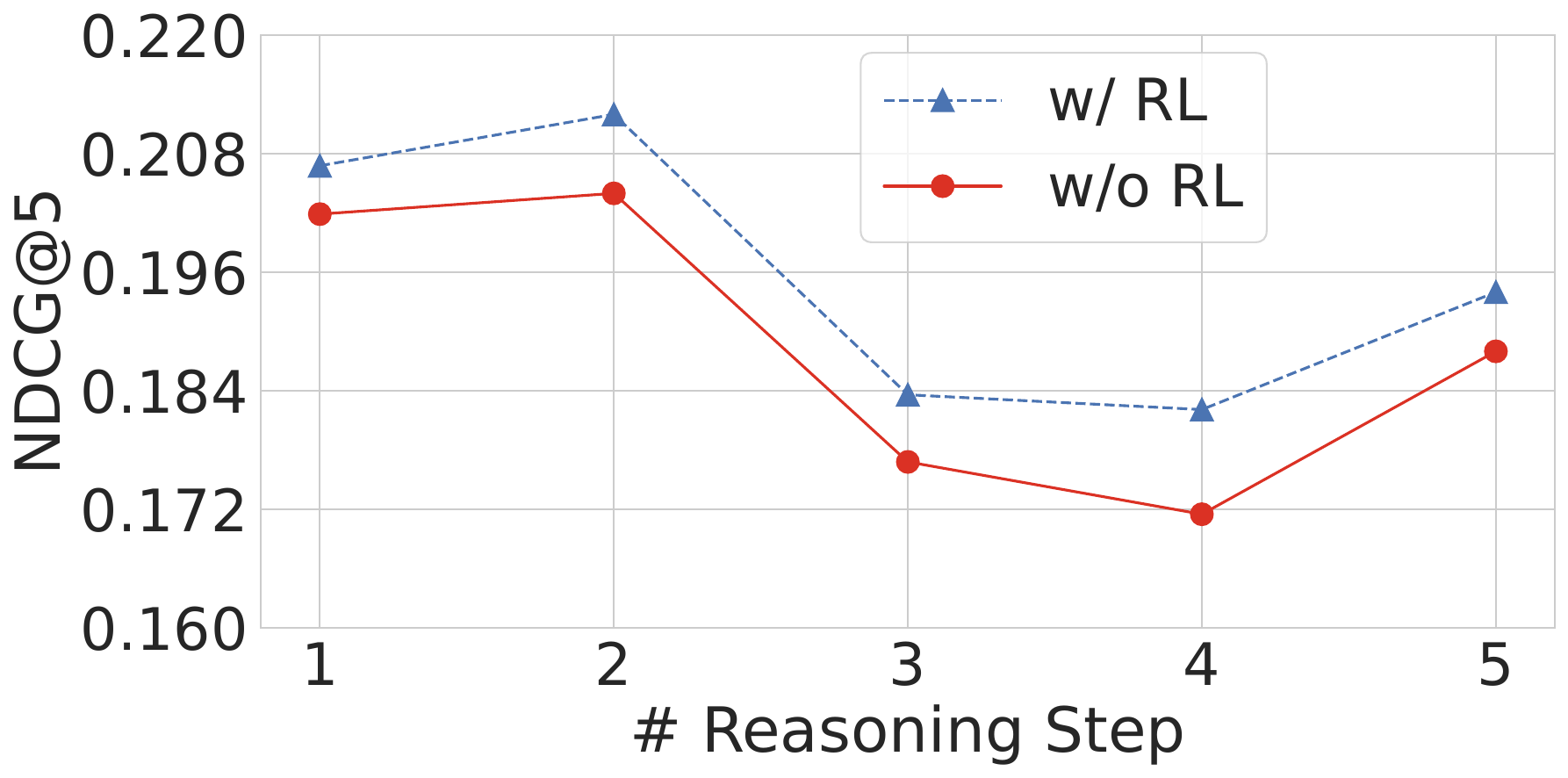}
 }
 \caption{
 Performance under different numbers of reasoning steps $K$, measured by NDCG@5 on the Qilin and KuaiSAR-Small datasets.
 }
 \label{fig:reason_step}
\end{figure}

\subsubsection{Impact of $\mathcal{L}_{\mathrm{TCL}}$}
To better understand the impact of $\mathcal{L}_{\mathrm{TCL}}$
(Eq.~\eqref{eq:target_cl}) on representation learning, we analyze the distance distribution between the hidden representation from the final reasoning step and the target item. Figure~\ref{fig:tcl_impact} compares the results with and without the loss term $\mathcal{L}_{\mathrm{TCL}}$.
We compare two settings: with MCA and without, where the latter uses only MSA during reasoning without accessing complementary historical information.
Without $\mathcal{L}_{\mathrm{TCL}}$, introducing MCA has little effect on the final distance. In contrast, when $\mathcal{L}_{\mathrm{TCL}}$ is applied, MCA produces representations that are closer to the target item, leading to more accurate predictions. This demonstrates that $\mathcal{L}_{\mathrm{TCL}}$ guides MCA to extract more target-relevant information.

\subsubsection{Impact of Reasoning Step}
We investigate the impact of reasoning steps in Figure~\ref{fig:reason_step}. On Qilin, performance generally improves with more steps, whereas on KuaiSAR-Small, excessive steps can hurt. This reflects varying reasoning depths across datasets—simpler samples may require fewer steps, and too many may cause ``overthinking'', similar to observations in LLMs~\cite{sui2025stop}. Thus, choosing an appropriate step number is crucial.
Additionally, RL consistently improves performance. Models with stronger pre-RL performance benefit more, indicating that both the base model's strength and RL's guidance are essential.

\begin{figure}[t]
 \centering
 \subfigure[Qilin]{
    \label{fig:qilin_tcl_w}
    \includegraphics[width=0.45\columnwidth]{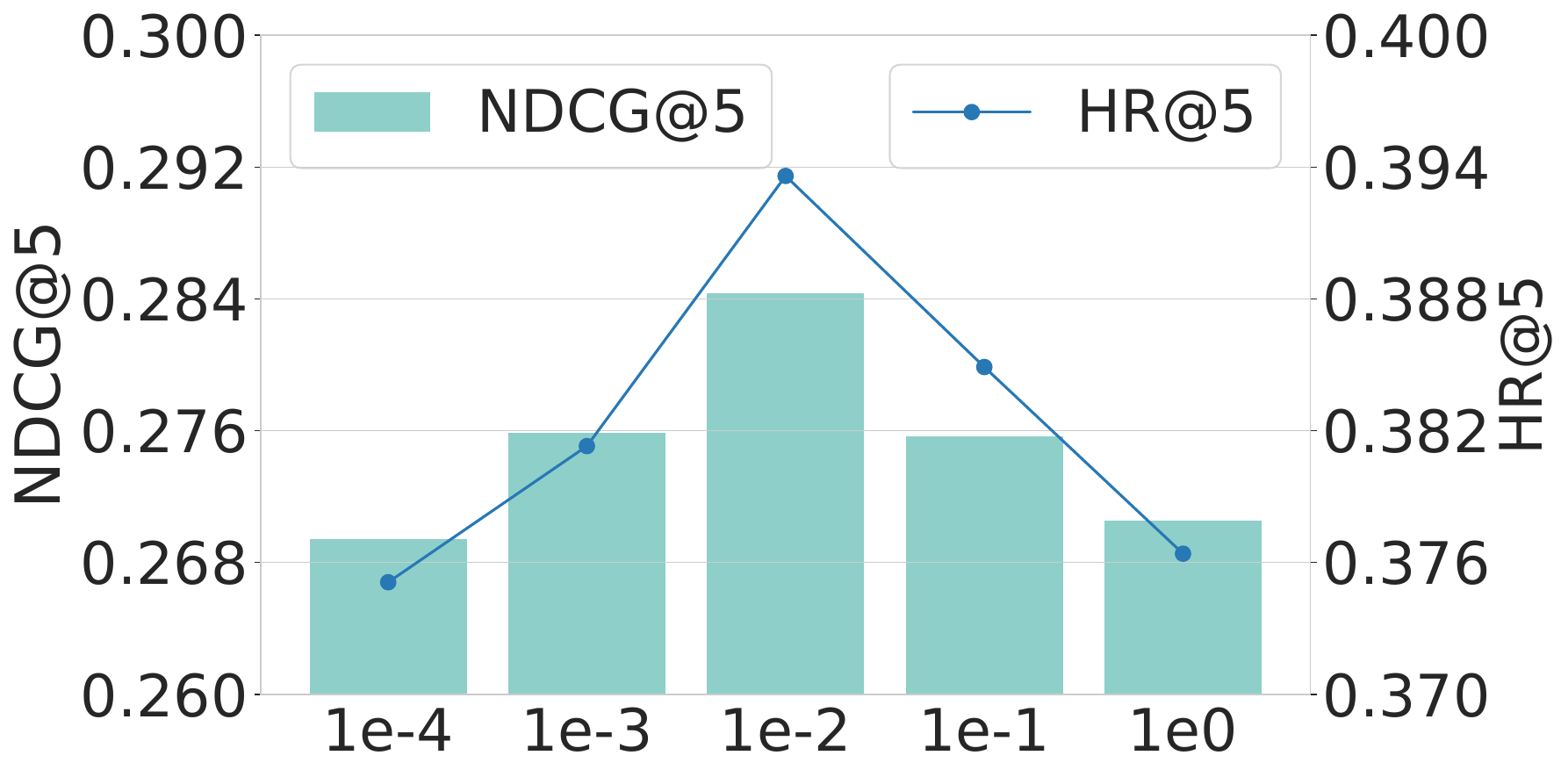}
 }
\subfigure[KuaiSAR-Small]{
    \label{fig:kuaisar_small_tcl_w}
    \includegraphics[width=0.45\columnwidth]{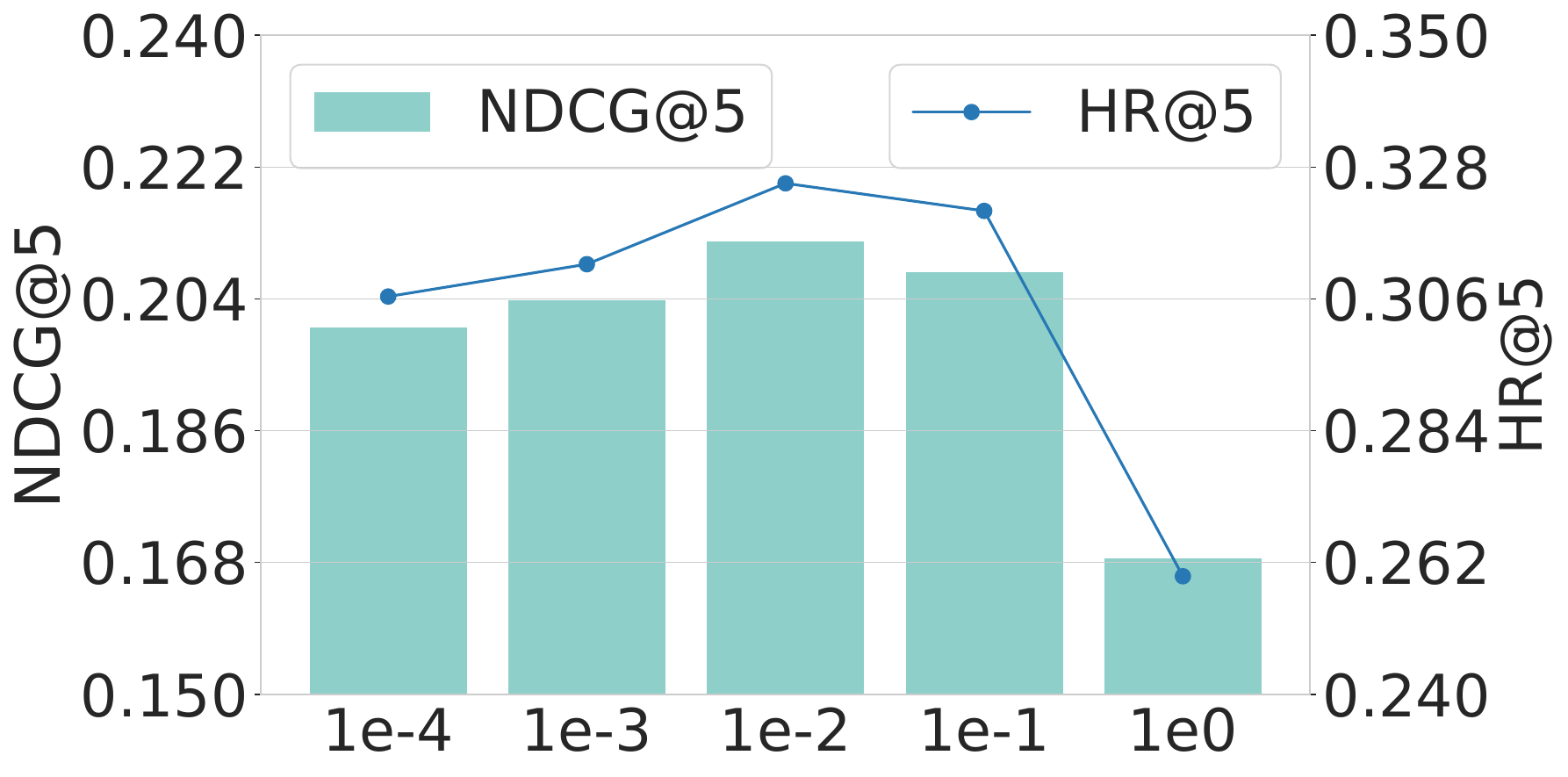}
 }
 \caption{
 Performance under different weights $\lambda_{\mathrm{TCL}}$ of the 
 loss $\mathcal{L}_{\mathrm{TCL}}$
 (Eq.~\eqref{eq:total_loss}), measured by NDCG@5 and HR@5 on the Qilin and KuaiSAR-Small datasets.
 }
 \label{fig:tcl_w}
\end{figure}

\begin{figure}[t]
 \centering
 \subfigure[Qilin]{
    \label{fig:qilin_kl_w}
    \includegraphics[width=0.45\columnwidth]{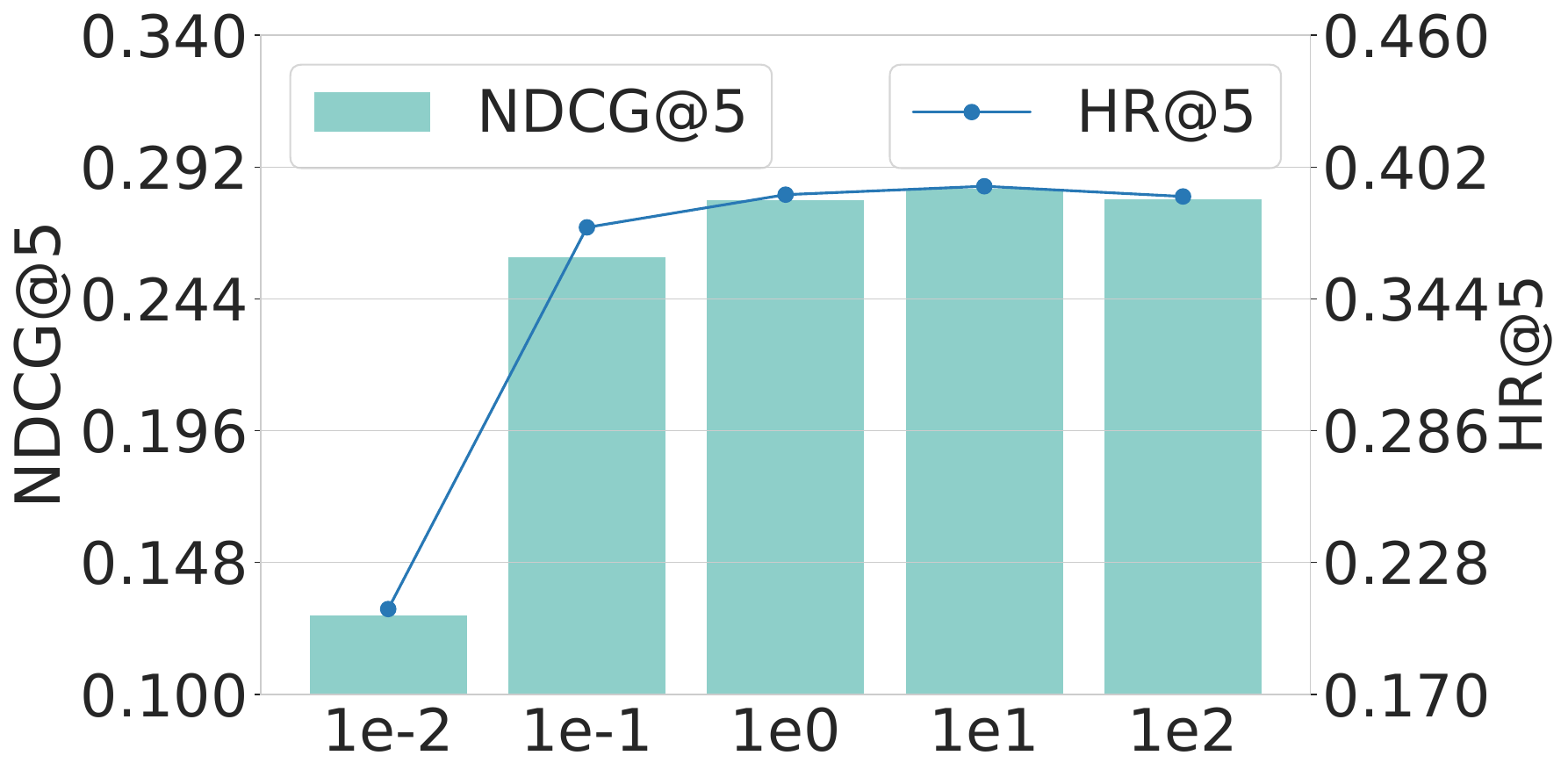}
 }
\subfigure[KuaiSAR-Small]{
    \label{fig:kuaisar_small_kl_w}
    \includegraphics[width=0.45\columnwidth]{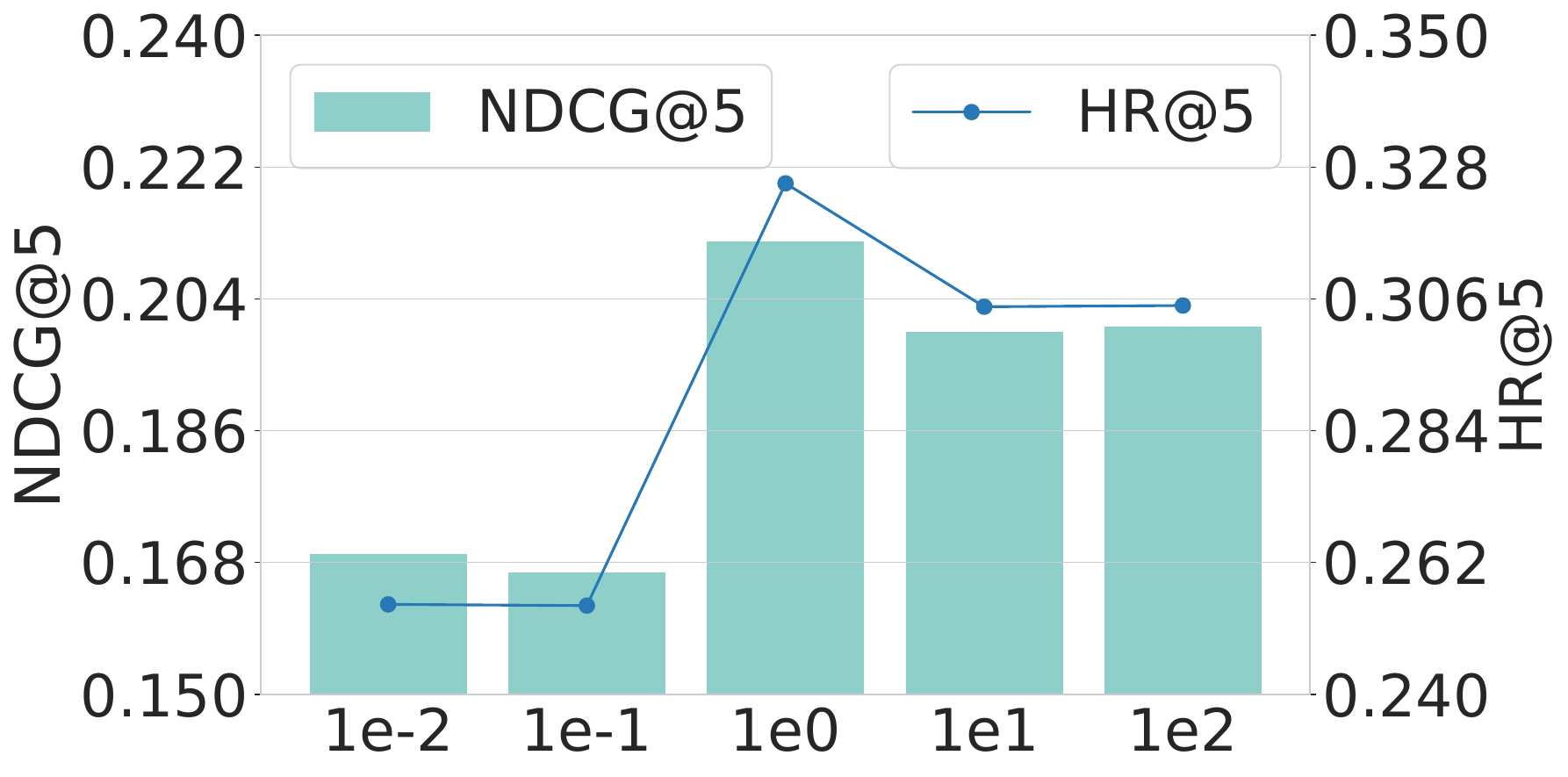}
 }
 \caption{
 Performance under different weights $\lambda_{\mathrm{KL}}$ of the KL divergence (Eq.~\eqref{eq:grpo}), measured by NDCG@5 and HR@5 on the Qilin and KuaiSAR-Small datasets.
 }
 \label{fig:kl_w}
\end{figure}

\subsubsection{Impact of Hyper-parameters}
We analyze the impact of key hyper-parameters on performance: the weight $\lambda_{\mathrm{TCL}}$ for the loss $\mathcal{L}_{\mathrm{TCL}}$ (Eq.~\eqref{eq:target_cl}) and the KL divergence weight $\lambda_{\mathrm{KL}}$ (Eq.~\eqref{eq:grpo}).
As shown in Figure~\ref{fig:tcl_w}, setting $\lambda_{\mathrm{TCL}}$ too high disrupts the optimization of the recommendation loss $\mathcal{L}_{\mathrm{rec}}$ (Eq.~\eqref{eq:total_loss}), while too low a value fails to guide the MCA module in capturing target-relevant information. These findings underscore the importance of appropriately tuning $\lambda_{\mathrm{TCL}}$.
Similarly, Figure~\ref{fig:kl_w} shows that excessively large and small values of $\lambda_{\mathrm{KL}}$ lead to suboptimal results. A large $\lambda_{\mathrm{KL}}$ limits the exploration capacity of the RL process, while a small value destabilizes training and weakens the model’s recommendation ability. These results highlight the importance of appropriately tuning $\lambda_{\mathrm{KL}}$.

\section{Conclusion}
This work addresses the challenge of search-enhanced recommendation by identifying a key limitation in existing methods: their inability to effectively extract search signals that are truly beneficial for recommendation, and vice versa. To overcome this, we propose \ourname, a novel framework based on latent cross reasoning. By iteratively reasoning over \srcandrec histories, \ourname selectively distills relevant signals from one behavior to enhance the other. In addition, reinforcement learning with GRPO encourages the model to explore diverse reasoning trajectories and optimize for recommendation performance. Experiments on public datasets demonstrate the effectiveness of \ourname.

\bibliographystyle{ACM-Reference-Format}
\bibliography{ref}

\end{document}